# Superconducting Cable Modelling into Electro-Magnetic Transient Simulation Tool

Christophe Creusot, Antonio Morandi *Senior Member, IEEE*, Francesco Mimmi, Emiliano Guerra, Alberto Bertinato, Pierre-Baptiste Steckler, Pier Luigi Ribani, Massimo Fabbri, Marco Bocchi, Andrea Musso, Giuliano Angeli, Diego Brasiliano

*Abstract* —The European project SCARLET aims to study and realize a demonstrator of a MVDC (Medium Voltage Direct Current) high-power superconducting cable. This device might be employed to connect offshore wind farms with land, expecting to significantly simplify the offshore platform by eliminating the need for its conversion function. For this purpose, windmill conversion chain must be modified to directly produce the MVDC export voltage.

In this scenario, this paper presents the case of a 1GW offshore windmill superconducting link and outlines the design consideration for a 1 GW onshore converter. For this cable, a protection strategy that combines DC circuit breakers with a Resistive Superconducting Fault Current Limiter is proposed. Moreover, this works demonstrates how a superconducting cable can be modelled as an electrical circuit to be integrated into a network simulation tool, enabling the investigation of various fault scenarios and protection strategies. Finally, a specific result is discussed to exemplify how the proposed approach can benefit the design of both the electrical network and the superconducting cable itself.

## I. INTRODUCTION

IN the framework of the European project SCARLET [1], the use of superconducting (SC) cables either onshore or offshore is being investigated for various application cases [2]. The main objective is to leverage the benefits of conducting large currents, typically exceeding 10 kA, while alleviating high voltage constraints by operating at medium DC voltage (typically below 100 kVdc). Specifically, two superconducting cable technologies are being designed and will be brought to the demonstration level: one utilizing High-Temperature Superconductor (HTS) tapes, operating at ±50 kVdc and cooled with liquid nitrogen, and the other using Magnesium Diboride ($MgB_2$) wires, operating at ± 25 kVdc and cooled with liquid hydrogen. The HTS superconducting cable will not only be designed for onshore applications but also for offshore use, considering a typical subsea length of 100 km. The $MgB_2$ cable is designed to be a hybrid energy transportation link combining electrical power transportation as well as hydrogen chemical energy transportation [3].

DC cable designs and constraints have been presented by different authors, mainly describing their architecture and their thermal and electric insulation systems [4]-[9]. The response to a fault has been addressed in [5]-[8]. A ±100 kV DC system including HTS power cable has been modelled in fault conditions in [7] and [8], considering a point-to-point and a multi-terminal layout respectively.

This article will describe the main technological components necessary to ensure the safe operation of the superconducting cable at medium voltage level, considering a 1 GW level of electrical power transmission. Additionally, it will discuss the electrical modeling of the superconducting cable and how it can be implemented into the commercial software EMTP® [10]-[11], which is specialized and commonly used for the simulation of transients in distribution or transmission grids, to evaluate the interaction between the DC system and the cable itself.

## II. REFERENCE CASE FOR 1 GW DC TRANSMISSION WITH SUPERCONDUCTING CABLES

One of the primary objectives of the SCARLET project is to demonstrate the feasibility of exporting bulk offshore wind power to onshore locations using a medium voltage DC superconducting cable. The main advantage of this approach is the ability to eliminate the offshore conversion platform. To achieve this, modifications are required to convert the existing AC output of the windmill into a DC output. This is feasible as the windmill already contains a conversion chain that can be adapted to produce the required DC voltage.

The greater challenge lies in transforming the onshore HVDC conversion station into an MVDC station. The current HVDC converters are limited in their nominal current ratings by individual components (IGBTs or IGCTs). In fact, due to the current constraints imposed by conventional cables, enhancing transmission power necessitates an elevation in voltage up to

The SCARLET project has received funding from the European Union's Horizon Europe research and innovation programme under grant agreement No. 101075602. (*Corresponding author: Christophe Creusot*).

C. Creusot, A. Bertinato, P.-B. Steckler and D. Brasiliano are with SuperGrid Institute, 69100 Villeurbanne, France (e-mail: christophe.creusot@supergrid-institute.com, alberto.bertinato@supergrid-institute.com, pierre-baptiste.steckler@supergrid-institute.com, diego.brasiliano@supergrid-institute.com).

A. Morandi, F. Mimmi, E. Guerra, P. L. Ribani and M. Fabbri are with the University of Bologna, Viale del Risorgimento 2, 40136 Bologna, Italy (e-mail: antonio.morandi@unibo.it, francesco.mimmi2@unibo.it, emiliano.guerra5@unibo.it, pierluigi.ribani@unibo.it, massimo.fabbri@unibo.it).

M. Bocchi, A. Musso and G. Angeli are with RSE S.p.A., Via Rubattino 54 20134 Milan, Italy (e-mail: marco.bocchi@rse-web.it, andrea.musso@rse-web.it, giuliano.angeli@rse-web.it)

525 kVdc in Europe [11]. The practice involves connecting a larger number of modules in series, effectively increasing the transmission voltage.

For the ±50 kVdc case, the proposed approach is to implement four Modular Multilevel Converters (MMC) in parallel, each of them fed by a 3-Phase transformer, as shown in Fig. 1. This setup naturally facilitates the converter protection by employing a DC breaker with a series reactor on the DC side and an AC breaker on the AC side. This configuration allows for the elimination of a fault in one converter while maintaining power transmission through the non-faulty converters.

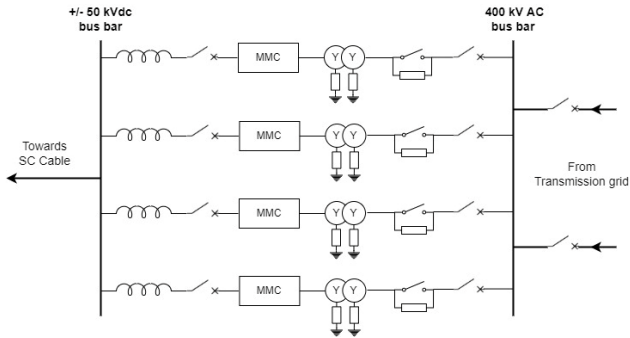

**Fig. 1.** Unifilar Onshore 1 GW ±50 kVdc converter station scheme.

DC circuit breakers are a subject of ongoing research in the HVDC field worldwide, with various manufacturers offering solutions. The solution presented in this study is a mechanical circuit breaker with active high-frequency current injection. It is based on 35 kV modules, so for the ±50 kVdc application, two modules are connected in series, while for the ±25 kVdc, only one module is necessary. Laboratory tests have demonstrated its capability to interrupt a current of up to 20 kA. Thanks to its electromagnetic actuator, it achieves a fault neutralization time of less than 5 ms [13].

The fast-acting DC breaker is advantageous for minimizing stress on the SC cable in case of a fault. Additionally, it is proposed to install a Resistive Superconducting Fault Current Limiter (RSFCL) in series with the SC cable. The RSFCL will absorb a significant amount of energy before fault neutralization by the DC circuit breakers. Its protective role becomes even more crucial in the event of DC breaker failure, where fault current interruption would rely on AC breakers, resulting in a short circuit duration on the order of 60 ms. Another significant advantage of the RSFCL is its ability to regenerate much faster than the SC cable in case it is not present and the fault results in a SC cable quench.

The schematic of the proposed 1 GW offshore export system is depicted in Fig. 2. Windmill clusters are of 200 MW size, they are connected to the DC busbar via medium voltage resistive cables. Possibly, one DC breaker can be installed at each cluster, allowing a continuous power flow in case a failure occurs in one cluster.

## III. CIRCUIT MODEL OF SUPERCONDUCTING CABLES

In this paper, an equivalent circuit model for the cable and its integration into EMTP® has been developed. This integration also encompasses the model of the DC system to evaluate their overall behavior, which is significantly influenced by their mutual interaction.

In the following, the model of the reference ±50 kVdc – 10 kA DC HTS cable is described in detail. The same modelling approach also applies to the ±25 kVdc – 20 kA DC MgB$_2$ cable. The layout of the reference cable is schematically shown in Fig. 3, it comprises a wired copper former, four HTS layers consisting of helically wound 2G tapes, and a stranded copper shield. Low-conductivity carbon black wrapping, indicated in black in Fig. 3, is included between all layers. Electrical insulation is placed between the outer HTS layer and the copper shield. The cable core is enclosed within a cryostat with two metallic pipe walls.

To develop the equivalent circuit, each of the physical conductors of the cable (HTS tapes, copper wires or tapes in the former and in the shield) is first individually modelled. A reduction procedure of the number of circuit components is later applied before implementing the circuit in EMTP® environment. The model of each individual conductor consists of a resistor in series with an inductor. All individual inductors are coupled with each other. The calculation of the induction coefficients is based on the precise geometrical model of the helical conductors, as described later. Metallic pipes of the cryostat are also included in the model by assuming that they only carry current in the longitudinal direction and by splitting them in a set of straight tile-shaped conductors parallel to the cable. In Fig. 4a, the electric circuit model for individual conductors is presented. This schematic depicts a cable section with a designated length ($L$), where each individual tape or wire is modelled as an insulated conductor. The electric circuit consists of clusters of parallel R-L branches representing all conductors for each cable's component (former, HTS layers, shield and pipes). The capacitive coupling between the outer HTS layer and the shield (through electric insulation), between the shield and the inner pipe (through coolant), and between the inner and outer pipe (through vacuum) is also included in the model.

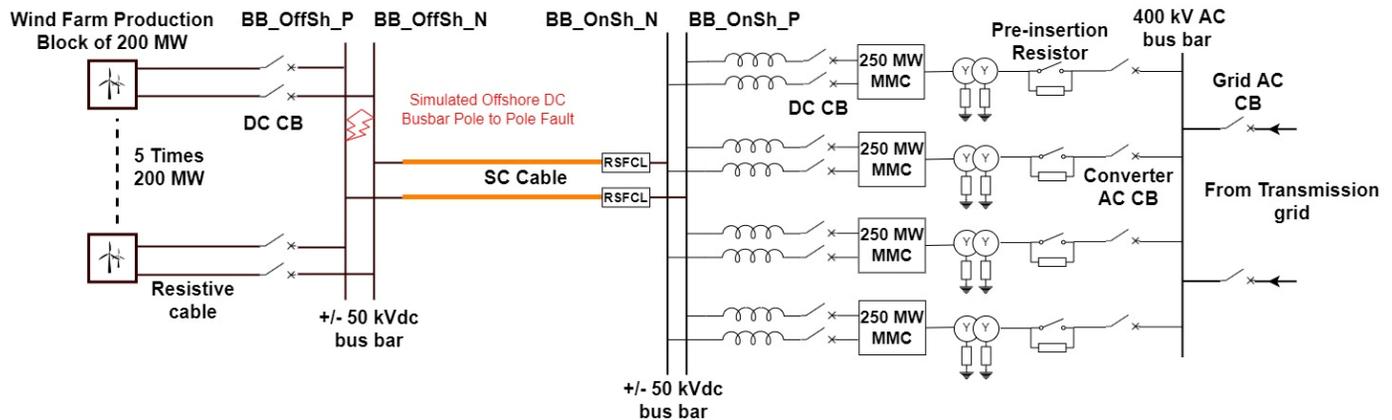

**Fig. 2.** Complete reference case schematic (DC bifilar and AC unifilar).

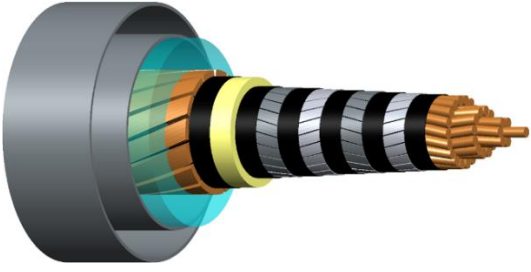

**Fig. 3.** Reference layout of the HTS cable.

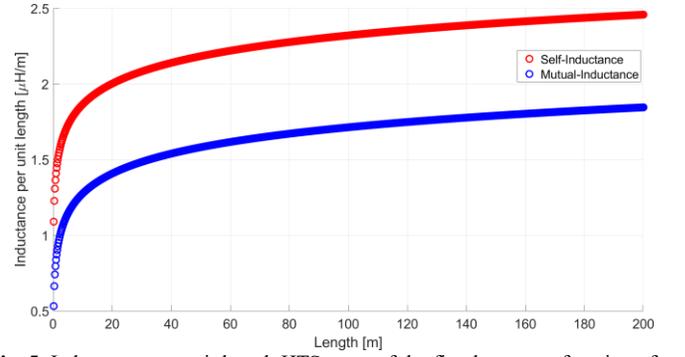

**Fig. 5.** Inductance per unit length HTS tapes of the first layer as a function of the axial length in the range 0.2 m (1 twist pitch) to 200 m (1000 twist pitches).

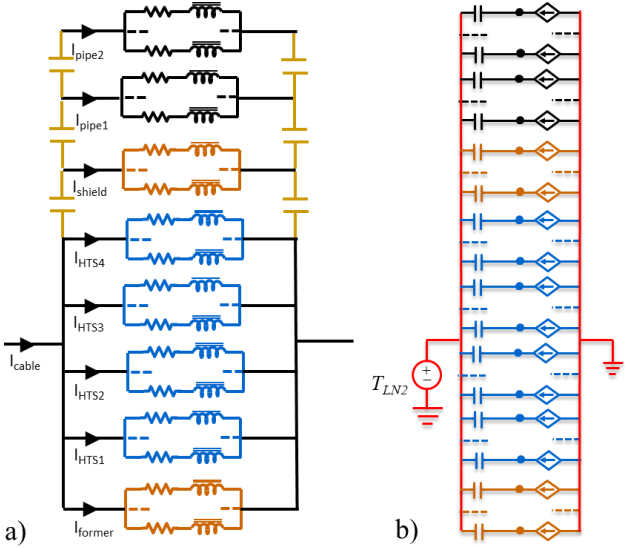

**Fig. 4.** Electro-thermal model for a section of the HTS cable, defined by its length ($L$), using individual conductor models. (a) Electric Circuit and (b) Thermal Network.

Since all circuit resistances are temperature-dependent, a coupled thermal network, shown in Fig. 4b, is introduced to calculate the time evolution of all conductors' temperatures based on the dissipation of the electric circuit. In the remainder of this Chapter, the parameters of the circuit model for individual conductors and the thermal network will be described in Sections A, B, and C. Finally, in Section D, the reduction procedure of the circuit is discussed.

### A. Inductances

The calculation of the mutual- and self-inductance coefficients of helically wound tapes (or wires) has been performed using a semi-analytical approach. More in particular, each of the tapes is divided into several thin filaments, whose axes are further subdivided into a number of straight segments. Grover's formula is employed to calculate the partial inductance between the two straight segments [14]. The final value of the inductance is obtained through a double summation over the filaments and the segments of the subdivision.

As an example, Fig. 5 shows the self-inductance per unit length of one HTS tape of the first layer as a function of the axial length of the helix, ranging from 1 twist pitch ($L_0 = 0.2$ m) to 1000 twist pitches (200 m). The presented inductances per unit length are normalised with respect to their corresponding axial length. The same plot also shows the mutual inductance per unit length between one HTS tape of the first layer and one HTS tape of the second layer displaced by 90 degrees with respect to the former one.

An exact logarithmic increase of the calculated inductances per unit length $m$ with respect to the length $L$ can be observed from Fig. 5, which can be interpolated by means of:

$$m = m_0 + k\ ln\left(\frac{L}{L_0}\right) \quad [\text{H/m}] \tag{1}$$

where $L_0$ is the twist pitch and $m_0$ and $k$ are fitting coefficients. Verification has been performed through numerical calculations to ensure that the logarithmic trend is strictly obeyed by any induction coefficient (self- or mutual-) of the conductors of the cable. This means that for calculating the self-induction coefficients $M$ of each cable conductors (or the mutual induction coefficients between pairs) of any length, only the fitting parameters $m_0$ and $k$ must be determined. Thus, the inductance $M$ can be obtained from the Eq. (2):

$$M = Lm = Lm_0 + k\ Lln\left(\frac{L}{L_0}\right) \quad [\text{H}] \tag{2}$$

To obtain the inductance coefficients, the self and mutual inductances per unit length of all the conductors of the cable were plotted against the axial length over an interval ranging from 1 to 10 twist pitches (that is, from 0.2 m to 2 m). Fitting parameters $m_0$ and $k$ were then obtained from these plots and used for calculating the inductance coefficients relative to the considered cable length via Eq. (2). Simplifying symmetry conditions were adopted (for example, the self-inductances of all tapes of a layer were considered to be identical, as well as mutual inductance between adjacent tapes in the same layer, etc…). For the reference cable layout of Fig. 3, 182 conductors are included in the model, consisting of all the copper wires of the former, all the tapes of the HTS layers, all the copper tapes of the shield, and all the straight tile-shaped conductors used for modelling the metallic pipes (each split in 24 straight conductors). The total calculation time of the full, symmetric, 182×182 inductances matrix was about 6 hours.

## B. Resistances and capacitances

Resistances of HTS tapes are obtained by modelling the superconductor by means of the well-known E-J power law, here stated in its inverse form:

$$J(E,T) = \frac{J_c(T)}{E_c}\left(\frac{E}{E_c}\right)^{\frac{1}{n}-1} E \quad [\text{A/m}^2]  \tag{3}$$

The following equivalent voltage- and temperature-dependant conductivity is defined based on Eq. (3)

$$\sigma_{HTS}(V,T) = \frac{J_c(T)}{E_c}\left(\frac{V}{L \cdot E_c}\right)^{\frac{1}{n}-1} + \sigma_{ns}(T) \quad [\text{S/m}]  \tag{4}$$

where the electric field $E$ can be substituted with the ratio $V/L$ and $\sigma_{ns}(T)$ is the normal state conductivity to which the power law must be smoothly connected during the transition [15],[16]. Based on the equivalent conductivity of the whole tape, defined in Eq. (5), a temperature and voltage- dependent resistance can be obtained for the HTS tape as in Eq. (6).

$$\sigma_{eq}(T) = \sigma_{Cu}(T)ff_{Cu} + \sigma_{Ag}(T)ff_{Ag} + \\ + \sigma_{HTS}(T)ff_{HTS} + \sigma_{Hast}(T)ff_{Hast}  \tag{5}$$

$$R(V,T) = \frac{L}{S_{tape}} \frac{1}{\sigma_{eq}(T)}  \tag{6}$$

where $S_{tape}$ is the cross-section of the tape, $\sigma_{Cu}$, $\sigma_{Ag}$, $\sigma_{HTS}$, $\sigma_{Hast}$ are the conductivities and $ff_{Cu}$, $ff_{Ag}$, $ff_{HTS}$, $ff_{Hast}$ are the filling factors of copper, silver, superconductor and Hastelloy within the tape. All other cable conductors (wires or tapes composing the former and the shield and straight conductors modelling the metallic pipes) are represented by considering temperature-dependent conductance using lookup tables. Capacitance of the equivalent circuit of Fig. 4, are calculated by means of the cylindrical capacitor formula. Capacitances are not considered between the HTS layers and between the former and HTS1 as no insulating medium is included in between.

## C. Thermal network

The temperatures of all conductors of the cable, required for the definition of the resistances of the electric circuit, are calculated by means of the thermal network shown in Fig. 4b. This is obtained by imposing an independent adiabatic thermal balance for each of the conductors in the form:

$$C_{eq}(T)\frac{dT}{dt} = \frac{V^2}{R(T,V)}  \tag{7}$$

where $C_{eq}$ is the equivalent thermal capacity of the conductors and, on the right side of Eq. (7), the dissipation term is implemented in the form of a controlled current source. The temperature of each conductor corresponds to the ground potential of the circuit nodes common to fictious capacitances and current source that are highlighted in Fig. 4b. For the purposes of the thermal network, the HTS tapes are modelled as a homogeneous material with an equivalent thermal capacity given by:

$$C_{tape}(T) = L \ S_{tape} \ \gamma_{tape} \ c_{tape}(T)  \tag{8}$$

where the mass density of the tape is defined as

$$\gamma_{tape} = ff_{Cu}\gamma_{Cu} + ff_{Ag}\gamma_{Ag} + ff_{HTS}\gamma_{HTS} + ff_{Hast}\gamma_{Hast}  \tag{9}$$

where $\gamma_{Cu}$, $\gamma_{Ag}$, $\gamma_{HTS}$, $\gamma_{Hast}$ represent are mass densities of copper, silver, superconductor and Hastelloy.
Moreover, the specific heat of the tape is computed as:

$$C_{tape}(T) = \\ \frac{ff_{Cu}\delta_{Cu}c_{Cu} + ff_{Ag}\delta_{Ag}c_{Ag} + ff_{HTS}\delta_{HTS}c_{HTS} + ff_{Hast}\delta_{Hast}c_{Hast}}{\gamma_{tape}}  \tag{10}$$

where $c_{Cu}$, $c_{Ag}$, $c_{HTS}$, $c_{Hast}$ are the temperature-dependent specific heats of the tape layers. It is pointed out that the adiabatic model represents a conservative assumption in terms of possible overtemperature that can arise in the cable during transient. However, heat exchange with the coolant and between the layers of the cable can be included in the model by adding transverse thermal conductances [17].

## D. Reduced equivalent circuit.

The equivalent circuit depicted in Fig. 4 comprises numerous branches, making it impractical for implementation in power system simulators. To make it suitable for integration into an Electro-Magnetic Transient modelling tool, a reduction procedure is applied to simplify its complexity, in conjunction with the model of the hosting DC system. The reduction procedure assumes that all conductors of the same cable layer (former, HTS layers, shield and pipes) carry the same current and operate at the same temperature. Therefore, resistive voltage drop is the same for all conductors of the same layer. Based on this assumption, all conductors of one layer can be merged into one unique conductor whereby the current follows a helical path, and the reduced cable model schematized in Fig. 6 is finally obtained. The corresponding equivalent thermal network is shown in Fig. 7.

Resistances, thermal capacitances and heat source terms of the reduced equivalent circuit of an individual layer are obtained from the parallel of the individual conductors belonging to the same layer. Based on magnetic conservation arguments or, equivalently, by equivalence of the overall circuit behaviour, the inductance coefficients of the reduced equivalent circuit are given by:

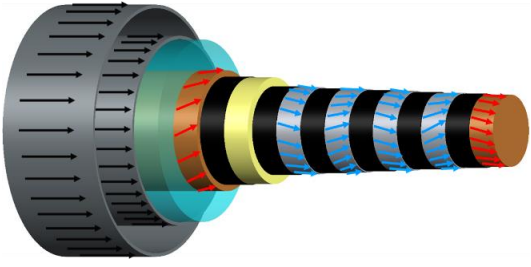

**Fig. 6.** Reduced model and current paths of the HTS power cable.

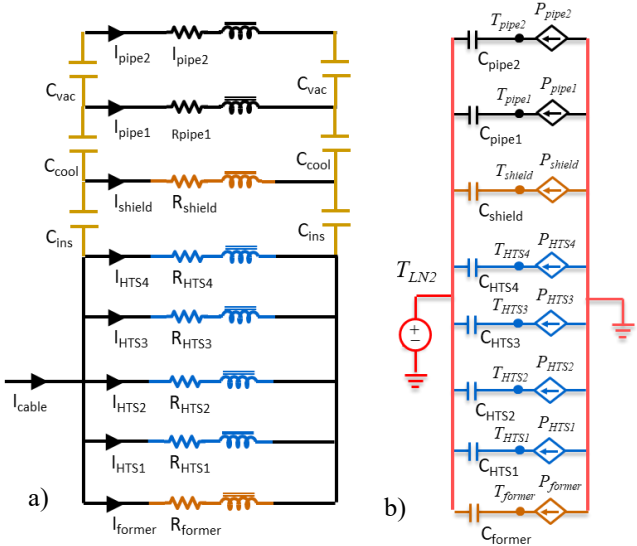

**Fig. 7.** Reduced equivalent circuits of the HTS power cable. (a) Electric Circuit and (b) Thermal Network.

$$M_{AB} = \frac{1}{n_A n_B} \sum_{i=1}^{n_A} \sum_{j=1}^{n_B} m_{ij} \qquad (11)$$

where $M_{AB}$ is the self/mutual inductance coefficient between two generic layers, $n_A$ and $n_B$ are the number of elements in layer A and B respectively, and $m_{ij}$ is the generic element of the complete inductance matrix of the cable. The reduced equivalent circuit consists of eight branches in total, corresponding to former, four HTS layer, shield and the two pipes respectively. It is worth noting that all parameters of the equivalent circuit are rigorously derived from the inherent physical properties of the material composing the conductor, such as electrical resistivity and specific heat capacity, both temperature-dependent. This approach eliminates the necessity for making *ad-hoc* assumptions.

## IV. RESULTS

To illustrate the implementation of the SC cable model into the full network model, the case of a pole-to-pole fault between the positive and the negative DC busbar of the wind-park side (labelled respectively as BB_Offsh_P and BB_Offsh_N in Fig. 2) is considered. For the illustration, the HTS cable is implemented in the ±50 kVdc system. The SC cable length is 100 km, the two SC cable poles are adjacent. All the components of the grid are modelled including the onshore converter as depicted in Fig. 1, that is described using the average arm model method [18]. The AC/DC converter is not an ideal DC source, therefore, using its actual model including its control and internal protection algorithm allows the computation of the real transient current when the fault occurs.

As the fault is located at 100 km from the onshore converter, the total inductance seen from the converter is very high, limiting the cable fault current to a low value. Also, as the initial power flow is from the left to the right of the circuit (the power is transferred from the wind park to the onshore converter station), and the AC grid feeds the fault through the converter station, the fault current flows in opposite direction to the steady state flow direction. Finally, because of the tripping of the onshore DC CBs, located on the DC side of the Onshore converter (Fig. 1), the cable current is interrupted after a few ms. As a result, the fault current starts from the steady state current of 10 kA and goes smoothly to zero without overshoot, this is shown in Fig. 8. In the example shown the fault is initiated at time 0,8 s, when the system is operating in normal condition transporting the rated current of 10 kA.

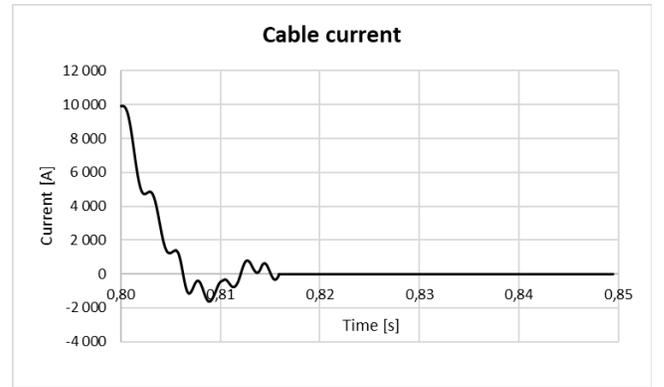

**Fig. 8.** Cable current during offshore DC busbar pole to pole fault.

Using the implemented cable model, it is possible to evaluate the currents in each cable layer (Fig. 9) and their corresponding temperature rises (Fig. 10). The following results can be highlighted:

a) Although the total fault current duration is in the order of 15 ms, current remains in the different layers, creating current loops with low resistive loss for approximately 500 ms.

b) After the cable current is cleared, there remains magnetic energy in the different layers that will dissipate in durations related to layer series resistances and direct inductance considering also mutual inductances that will still dictate current repartition before reaching a new steady state i.e. no current in any of the parallel branches.

c) During the fault, there is an uneven current sharing between the HTS layers. This distribution is driven by the mutual inductances.

d) The copper former is parallel to the HTS layers with no coupling capacitance. In contrary, the copper shield is parallel to the HTS layers but with coupling capacitances and grounding at its two extremities. It results in a current flow in the copper former in the same direction as the cable current and, contrary, a current flow in the copper shield in the opposite direction to the cable current.

e) HTS4 layer current flows in the opposite direction to the cable current and the copper shield current.

f) In HTS2 layer, the current reaches a peak value of 4 kA, briefly exceeding its critical current of 3000 A.

g) The most stressed layers in terms of current peak and duration are the former and the shield.

The copper former and shield are subjected to a relatively high current and extended current duration, which might lead to overheating. However, as depicted in the temperature plot shown in Fig. 10, the temperature rise of the shield is not significant, and the temperature increase in the HTS layers is on the order of 0.1 K, thus not affecting the cable's integrity.

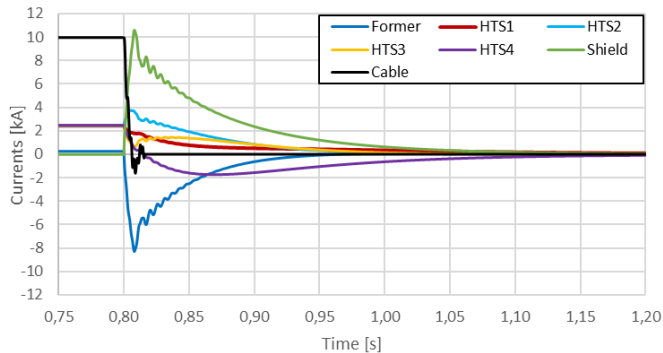

**Fig. 9.** Current distribution in the cable conductive layers.

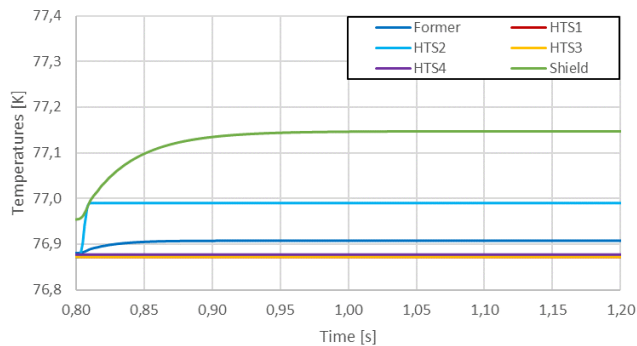

**Fig. 10.** Temperature distribution in the cable conductive layers.

This example shows that, although the cable fault behaviour appears to be non-critical in terms of current peak and current duration, the response of each individual layer is not so obvious. The investigation of each individual current and temperature is required to ensure there is no risk for the cable safety. It is also stressed that using simultaneously an accurate model of both the superconducting cable and the hosting power system, which mutually impact each other, is required in order to draw reliable conclusion concerning their overall behaviour during the fault. Due to its widespread application for modelling power systems, the EMTP® software was chosen for carrying out the transient analysis and the circuit model of the superinducing cable was developed therein too.

## V. Conclusion

In this paper, a methodology for modelling the equivalent electrical circuit of superconducting cables is presented. This cable model, applied to the ±50 kVdc HTS cable, is implemented into the full 1 GW offshore windfarm export electrical system where the onshore converter is described as a real component using the average arm model and RSFCL as well as DC circuit breakers are modelled. One fault case is simulated, and its results are discussed. It illustrates how the cable interacts with the full electrical system and how the cable internal behaviour can be evaluated in case of a transient current event. Based on these results, the SCARLET project will proceed with a simulation plan that encompasses both HTS and MgB₂ cable cases in order to properly design the protection devices, including the RSFCL as well as the cables themselves.